\PassOptionsToPackage{unicode}{hyperref}
\PassOptionsToPackage{hyphens}{url}
\documentclass[12pt]{article}

\usepackage{xcolor}
\usepackage{amsmath,amssymb}
\usepackage{graphicx}
\usepackage{longtable,booktabs,array}
\usepackage{hyperref}

\usepackage[T1]{fontenc}
\usepackage[utf8]{inputenc}
\usepackage{lmodern}

\usepackage[margin=1in]{geometry}
\hypersetup{
  hidelinks,
  pdfcreator={LaTeX via pandoc}}

\begin{document}

\title{\textbf{Beyond Calibration: Confounding Pathology Limits Foundation Model Specificity in Abdominal Trauma CT}}

\author{
Jineel H Raythatha BBmed, MD, MS\textsuperscript{1,2,3}, Shuchang Ye BAC\textsuperscript{1}, \\
Jeremy Hsu BPharm, MBBS, DClinSurg, FRACS, FACS\textsuperscript{2,3}, \\
Jinman Kim BCST, PhD\textsuperscript{1}
}

\date{}

\maketitle

\noindent\textbf{Affiliations:}

\noindent\textsuperscript{1}School of Computer Science, Faculty of Engineering, The University of Sydney \\[0.5ex]
\textsuperscript{2}Department of Surgery, Faculty of Medicine and Health, The University of Sydney \\[0.5ex]
\textsuperscript{3}Westmead Trauma Service, Westmead Hospital, Western Sydney Local Health District

\vspace{0.3cm}
\noindent\hrule
\vspace{0.25cm}

\noindent\textbf{Summary statement:} Foundation models matched task-specific models for traumatic bowel injury detection without training, but their specificity dropped 50\% points in the presence of confounding solid organ injuries versus 12\% points for a supervised ensemble, driven by negative-class heterogeneity.

\vspace{0.3cm}
\noindent\hrule
\vspace{0.3cm}

\noindent\textbf{Key Points:}
\vspace{0.1cm}

\begin{enumerate}
\setlength{\itemsep}{0.1cm}
\item When negative cases included concurrent solid organ injuries, foundation model specificity fell by 50--51 percentage points compared with 12--41 percentage points for task-specific models; because both subgroups had zero bowel injury prevalence, this difference reflects confounding pathology rather than prevalence miscalibration.
\item Foundation models achieved discrimination equivalent to task-specific models (area under the curve 0.64--0.68 versus 0.58--0.64) for traumatic bowel injury detection without any task-specific training, though with higher sensitivity (79--91\%) at the cost of lower specificity (33--50\%).
\item Comparing specificity across negative subgroups with controlled target-class prevalence provides a general framework for diagnosing whether specificity deficits stem from prevalence miscalibration or confounding pathology, applicable beyond bowel injury detection.
\end{enumerate}

\vspace{0.3cm}
\noindent\hrule
\vspace{0.3cm}

\section{Abstract}

\textbf{Purpose:} Translating foundation models into clinical practice requires evaluating their performance under compound distribution shift, where severe class imbalance coexists with heterogeneous imaging appearances. This challenge is particularly relevant for traumatic bowel injury, a rare but high-mortality diagnosis. We investigated whether specificity deficits in foundation models are associated with heterogeneity in the negative class.

\textbf{Methods:} This retrospective study used the multi-institutional, international, RSNA Abdominal Traumatic Injury CT dataset (2019-2023), comprising scans from 23 centres. Two foundation models (MedCLIP, zero-shot; RadDINO, linear probe) were compared against three task-specific approaches (CNN, Transformer, Ensemble). Models were trained on 3,147 patients (2.3\% bowel injury prevalence) and evaluated on an enriched 100-patient test set. To isolate negative-class effects, specificity was assessed separately in patients without bowel injury who had concurrent solid organ injury (n=58) versus no abdominal pathology (n=50).

\textbf{Results:} Foundation models achieved equivalent discrimination to task-specific models (AUC, 0.64--0.68 versus 0.58--0.64) with higher sensitivity (79--91\% vs 41--74\%) but lower specificity (33--50\% versus 50--88\%). All models demonstrated high specificity in patients without abdominal pathology (84--100\%). However, when solid organ injuries were present, specificity declined substantially for foundation models (50--51\% points) compared with smaller reductions 12--41\% points for task-specific models.

\textbf{Conclusion}: Foundation models matched task-specific discrimination without task-specific training, but their specificity deficits were driven primarily by confounding negative-class heterogeneity rather than prevalence alone. Susceptibility to this negative-class heterogeneity decreased progressively with labelled training, suggesting that targeted adaptation is required before clinical implementation.

\textbf{Keywords:} Foundation models, distribution shift, traumatic bowel injury, class imbalance, computed tomography, deep learning

\clearpage

\section{Introduction}

Artificial intelligence in radiology has expanded rapidly, with more than 900 FDA-cleared algorithms and growing adoption in surgical practice \cite{sosna2025,sivakumar2025}. Yet model performance under distribution shift --- where test conditions differ from training --- remains a critical barrier to reliable deployment \cite{godau2025,nguyen2023}. This challenge is particularly acute for pathologies characterised by compound distribution shift -- the intersection of class imbalance and appearance heterogeneity. Current AI techniques addressing class imbalance assume consistent appearance within each class, whereas techniques addressing appearance heterogeneity assume sufficient training examples \cite{nguyen2023,qu2020}. When both co-occur, a common scenario in clinical practice, neither approach is sufficient.

Traumatic bowel injury exemplifies compound distribution shift. With prevalence of 2--5\% in blunt abdominal trauma \cite{fakhry2003}, it represents severe class imbalance. Unlike conditions with consistent appearance signatures, bowel injury manifests with remarkable appearance heterogeneity: from subtle mesenteric stranding to frank pneumoperitoneum, from focal wall thickening to diffuse oedema, across varying anatomical locations \cite{butela2001}. Delayed or missed diagnosis increases mortality significantly, making accurate detection critical for surgical planning and operative decision-making \cite{fakhry2003}. This combination of class imbalance, appearance heterogeneity, and clinical consequence makes bowel injury an ideal testbed for evaluating model behaviour under compound distribution shift.

Medical foundation models -- large-scale neural networks pre-trained on diverse medical imaging datasets -- theoretically offer advantages for compound distribution shifts \cite{nguyen2023,moor2023}. Pre-training on heterogeneous data creates feature representations robust to appearance variation \cite{moor2023}. Vision-language models like CLIP and its medical variants go further, aligning visual and textual representations to enable zero-shot classification without task-specific training \cite{silva2025,radford2021}. Zero-shot classification is particularly appealing for rare pathologies where labelled training data are scarce. Yet whether these advantages persist when class imbalance and appearance heterogeneity co-occur remains uncharacterised. Performance deficit in compound distribution shift is typically attributed to prevalence miscalibration --- the mismatch between balanced pre-training distributions and imbalanced deployment conditions \cite{godau2025}. However, in trauma imaging, an alternative mechanism may contribute: the negative class is itself heterogeneous, as patients without bowel injury frequently have concurrent solid organ injuries with overlapping imaging features \cite{bates2017}. Whether prevalence miscalibration or this confounding concurrent pathology is the primary driver has not been examined.

This study compares two foundation models against three task-specific approaches for traumatic bowel injury detection representing compound distribution shift. With large-scale, diverse pre-training, we hypothesised foundation models would achieve equal discrimination for bowel injury compared to task-specific approaches. However, given that foundation models may not have been trained on rare pathologies such as bowel injury, we hypothesised that they may exhibit specificity deficits. To test this, we performed stratified analysis comparing model specificity on patients with concurrent solid organ injuries against patients without abdominal pathology.

\clearpage

\section{Methods}

\subsection{Study Design}

\begin{figure}[htbp]
    \centering
    \includegraphics[width=0.95\textwidth]{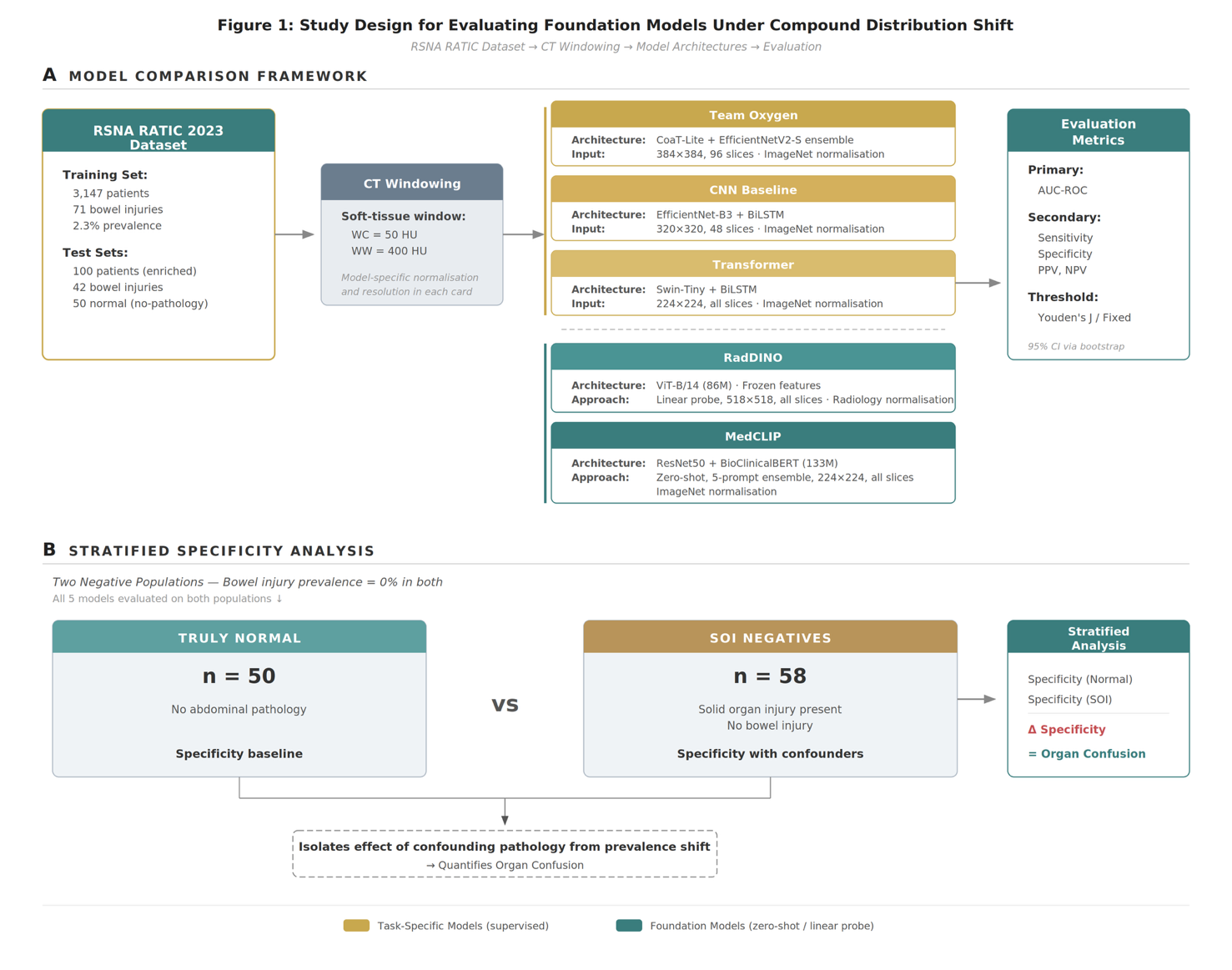}
    \caption{Study Design for Evaluating Foundation Models Under Compound Distribution Shift. Five models were evaluated: two foundation models (MedCLIP, zero-shot; RadDINO, linear probe) and three task-specific approaches (CNN Baseline, Transformer, Team Oxygen). The training set (n = 3,147; 2.3\% bowel injury prevalence) reflects natural prevalence. The enriched test set (n = 100; 42\% prevalence) enables statistical evaluation. The no-pathology test set (n = 50; 0\% prevalence, no abdominal pathology) enables stratified specificity analysis.}
    \label{fig:study-design}
\end{figure}

We conducted a retrospective evaluation comparing two foundation models against three task-specific approaches for bowel injury detection on computed tomography (CT) (Figure \ref{fig:study-design}). This study used publicly available, de-identified data from the RSNA RATIC 2024 Challenge; institutional review board approval was not required.

\subsection{Dataset}

We used the RSNA Abdominal Traumatic Injury CT (RATIC) dataset \cite{rudie2024}, comprising contrast-enhanced abdominal CT scans from trauma patients across 23 institutions internationally. The training set contained 3,147 patients, of whom 71 (2.3\%) had bowel injury. Training distributions were not modified. Table \ref{tab:dataset} presents full dataset characteristics.

\begin{table}[htbp]
\centering
\caption{Dataset characteristics}
\label{tab:dataset}
\begin{tabular}{@{}lccc@{}}
\toprule
\textbf{Characteristic} & \textbf{Training Set} & \textbf{Enriched Test Set} & \textbf{No-pathology Test Set} \\
\midrule
Total patients & 3,147 & 100 & 50 \\
Bowel injury & 71 (2.3\%) & 42 (42.0\%) & 0 (0\%) \\
Age, mean $\pm$ SD (y) & 46.6 $\pm$ 20.0 & 40.1 $\pm$ 17.5 & 53.0 $\pm$ 22 \\
Male sex & 2,164 (68.8\%) & 75 (75.0\%) & 24 (48.0\%) \\
\bottomrule
\end{tabular}
\end{table}

An enriched test set of 100 patients (42 with bowel injury) was constructed separately from training data to enable statistical evaluation across injury presentations. The test set was younger (mean 40.1 vs 46.6 years) with a higher proportion of males (75.0\% vs 68.8\%) than the training set. 58 had concurrent solid organ injuries (liver, spleen, or kidney).

A separate no-pathology cohort of 50 patients without abdominal pathology was constructed for stratified specificity analysis from the training data. Comparing model performance on this cohort against the 58 patients with solid organ injuries but no bowel injury --- both groups having zero bowel injury prevalence --- isolates the effect of negative-class composition on specificity.

\subsection{Models}

Five models were evaluated (Table \ref{tab:models}): two foundation models and three task specific approaches. All models used the same data for training and testing unless specified.

\begin{table}[htbp]
\centering
\caption{Model specifications}
\label{tab:models}
\footnotesize
\begin{tabular}{@{}llllll@{}}
\toprule
\textbf{Model} & \textbf{Architecture} & \textbf{Input Size} & \textbf{Approach} & \textbf{Training} & \textbf{Params} \\
\midrule
\multicolumn{6}{l}{\textit{Foundation models}} \\
MedCLIP & ResNet50 + & 224×224, & Zero-shot, & None & $\sim$133M \\
& BioClinicalBERT & all slices & 5-query ensemble & (pre-trained) & \\
RadDINO & ViT-B/14 & 518×518, & Linear probe, & L2 logistic regression, & $\sim$86M \\
& & all slices & frozen features & 5-fold CV & \\
\midrule
\multicolumn{6}{l}{\textit{Task-specific models}} \\
CNN Baseline & EfficientNet-B3 + BiLSTM & 320×320, & 2.5D sequence & Focal loss, AdamW & $\sim$15M \\
& & 48 slices & & & \\
Team Oxygen & CoaT-Lite + EffNetV2-S & 384×384, & 2.5D ensemble & 4-fold CV, AdamW & $\sim$200M \\
& + LSTM/GRU & 96 slices & & & \\
Transformer & Swin-Tiny + Linear & 224×224, & Two-stage, & BCE loss, AdamW & $\sim$28M \\
& & all slices & mean aggregation & & \\
\bottomrule
\end{tabular}
\end{table}

\subsubsection{MedCLIP Foundation Model}

MedCLIP is a vision-language foundation model adapted from CLIP for medical imaging \cite{wang2022}. We selected MedCLIP because it is well-established, publicly available, and specifically adapted for medical imaging with clinical text alignment. Its zero-shot inference capability provides the clearest test of foundation model behaviour under compound distribution shift, as no task-specific training is involved.

It comprises a ResNet50 image encoder (25M parameters) and Bio\_ClinicalBERT text encoder (110M parameters) \cite{wang2022}, projecting both modalities to a shared 256-dimensional embedding space through contrastive pre-training on 217,000 medical image-text pairs.

Zero-shot inference was employed without task-specific fine-tuning. Prompt engineering followed systematic evaluation across five categories: anatomical descriptions, radiological findings, clinical terminology, combined descriptors, and negation patterns. A five-prompt ensemble with majority voting (threshold 0.8, requiring at least four of five prompts to agree) was selected based on validation performance following established prompting strategy \cite{sivarajkumar2023}.

For inference, all CT slices per patient were processed at 224×224 pixels with soft-tissue windowing (width 400, level 50 HU) and greyscale-to-RGB conversion. Maximum aggregation across slices determined patient-level predictions.

\subsubsection{RadDINO Foundation Model}

We used RadDINO as a feature extractor with a logistic regression classifier \cite{perez2025}. Patient-level embeddings were computed by extracting CLS token features from each CT slice and applying maximum pooling across slices. Features were standardised to zero mean and unit variance. A logistic regression classifier with L2 regularisation was trained on these frozen features, with the regularisation parameter (C) selected via five-fold stratified cross-validation optimising Area under the receiver operating characteristics curve (AUC). Class imbalance was addressed using inverse-frequency weighting.

For inference, CT slices were processed at 518×518 pixels (RadDINO's native resolution) with identical soft-tissue windowing to MedCLIP (width 400, level 50 HU) but radiology-specific normalisation (mean=0.531, std=0.258) rather than ImageNet normalisation. Maximum aggregation across slices determined patient-level predictions. The operating threshold was determined using Youden's J statistic on cross-validation predictions.

\subsubsection{CNN Baseline}

We developed a 2.5D CNN using EfficientNet-B3 (ImageNet pre-trained) with bidirectional long short-term memory (LSTM) for slice aggregation. CT images were windowed (width 400, level 50 HU) and resized to 320×320 pixels. Up to 48 slices were uniformly sampled per patient. Class imbalance was addressed through focal loss ($\alpha$=0.75, $\gamma$=3.0), 15× oversampling of positive cases, and 3× loss weighting for bowel predictions. Training used AdamW optimiser (learning rate 1×10$^{-4}$, weight decay 1×10$^{-5}$), batch size 2 with 8× gradient accumulation, and early stopping.

\subsubsection{Transformer Baseline}

We implemented a Swin Transformer with bidirectional LSTM for temporal aggregation. A two-phase training strategy was employed: phase one trained on balanced slice-level samples (1,500 injury slices, 1,500 healthy slices) with cross-entropy loss; phase two added multi-organ learning with 100× bowel loss weighting. Training used AdamW with learning rate 2×10$^{-4}$. Patient-level predictions used mean aggregation across slice probabilities.

\subsubsection{Team Oxygen}

Team Oxygen achieved first place in the RSNA RATIC 2024 competition trained with the same data. Their solution employed an ensemble of 2.5D models combining convolutional neural networks (CoaT-Lite Medium, CoaT-Lite Small, EfficientNetV2-S) with recurrent layers (Gated Recurrent Unit and LSTM). Training used binary cross-entropy combined with Dice loss, AdamW optimiser, and four-fold cross-validation. We retrained their publicly available implementation; 10 of 14 classification models were successfully reproduced. The remaining 4 models did not have available weights and were related to extravasation which is not assessed in this study.

\subsection{Evaluation}

AUC served as the primary metric given its threshold-independence, enabling fair comparison between zero-shot inference (fixed threshold) and task-specific models (optimised thresholds). Secondary metrics included F1 score, sensitivity, specificity, and positive predictive value (PPV) at each model's operating point. Operating thresholds for task-specific models were determined using Youden's J statistic; MedCLIP used the fixed majority voting threshold. Confidence intervals (95\%) were computed via bootstrap resampling (1,000 iterations).

Stratified specificity analysis compared model performance on two negative subgroups: patients with no abdominal pathology (n=50) and patients with solid organ injuries but no bowel injury (n=58). Both groups had zero bowel injury prevalence, isolating the effect of confounding pathology on model specificity.

\clearpage

\section{Results}

\subsection{Model Performance}

\begin{figure}[htbp]
    \centering
    \includegraphics[width=0.95\textwidth]{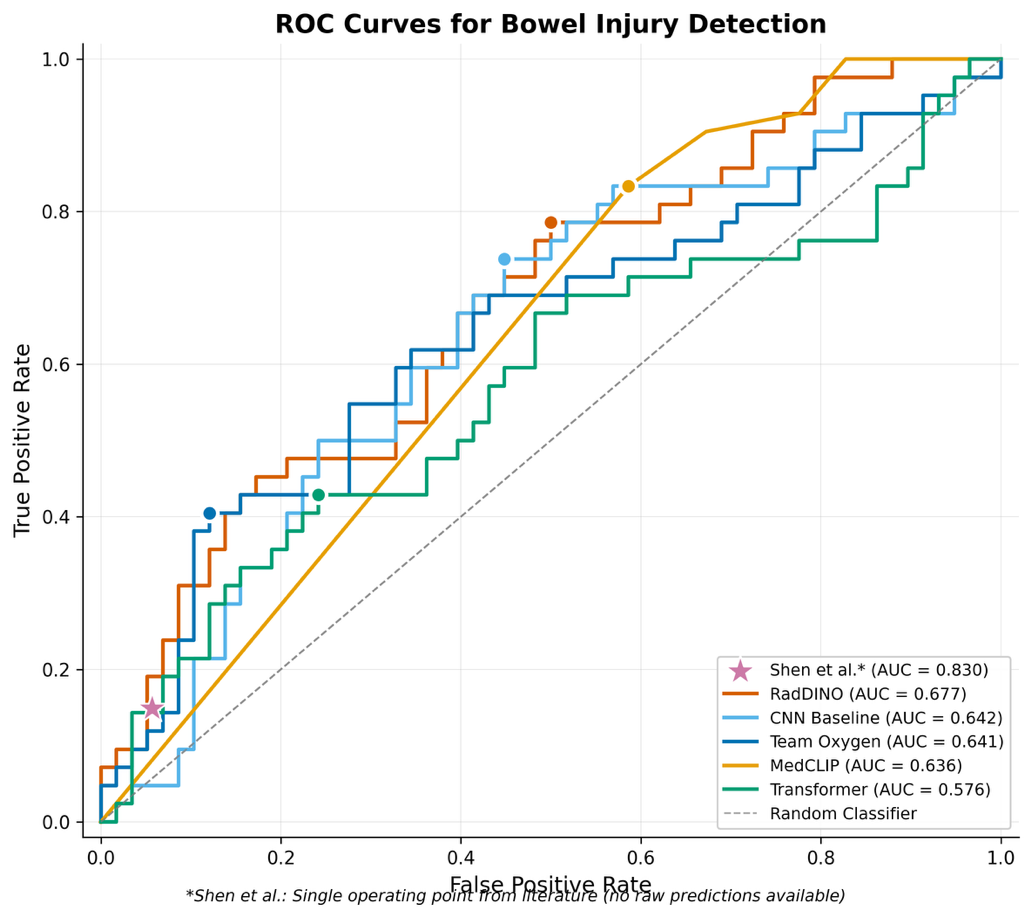}
    \caption{ROC curves showing discrimination performance for each model. Operating points indicated by markers. Foundation models shown in warm colours (RadDINO, orange; MedCLIP, gold); task-specific models in cool colours (CNN Baseline, light blue; Team Oxygen, dark blue; Transformer, green). The star indicates Shen et al.'s published operating point from cross-validation (raw predictions unavailable for curve generation).}
    \label{fig:roc}
\end{figure}

Table \ref{tab:performance} and Figure \ref{fig:roc} present model performance on the enriched test set. Foundation models achieved equivalent discrimination to task-specific models (AUC 0.64--0.68 versus 0.58--0.64; Table \ref{tab:performance}). However, operating characteristics differed markedly. Foundation models favoured sensitivity over specificity --- MedCLIP achieved 90.5\% sensitivity but only 32.8\% specificity --- while task-specific models showed the inverse pattern, with Team Oxygen reaching 87.9\% specificity but only 40.5\% sensitivity. RadDINO and the CNN baseline fell between these extremes. Full confidence intervals are reported in Table \ref{tab:performance}.

\begin{table}[htbp]
\centering
\caption{Model performance on enriched test set (n=100, 42\% bowel injury prevalence)}
\label{tab:performance}
\footnotesize
\begin{tabular}{@{}lcccccc@{}}
\toprule
\textbf{Model} & \textbf{AUC [95\% CI]} & \textbf{Sensitivity, \%} & \textbf{Specificity, \%} & \textbf{F1 [95\% CI]} & \textbf{PPV} & \textbf{NPV} \\
\midrule
\multicolumn{7}{@{}l}{\textit{Foundation models}} \\
RadDINO & 0.68 [0.57--0.79] & 78.6 [64.9--89.7] & 50.0 [36.2--62.7] & 0.63 [0.52--0.73] & 0.53 & 0.76 \\
MedCLIP & 0.64 [0.55--0.73] & 90.5 [80.0--97.9] & 32.8 [21.4--45.5] & 0.64 [0.53--0.74] & 0.49 & 0.83 \\
\midrule
\multicolumn{7}{@{}l}{\textit{Task-specific models}} \\
CNN Baseline & 0.64 [0.53--0.74] & 73.8 [60.5--86.1] & 55.2 [42.1--66.7] & 0.63 [0.51--0.73] & 0.54 & 0.74 \\
Team Oxygen & 0.64 [0.53--0.75] & 40.5 [25.6--54.5] & 87.9 [79.2--95.9] & 0.52 [0.36--0.65] & 0.71 & 0.67 \\
Transformer & 0.58 [0.46--0.70] & 42.9 [28.2--57.9] & 75.9 [64.7--86.2] & 0.49 [0.33--0.62] & 0.56 & 0.65 \\
\bottomrule
\end{tabular}
\end{table}

\subsection{Stratified Specificity Analysis}

\begin{figure}[htbp]
    \centering
    \includegraphics[width=0.95\textwidth]{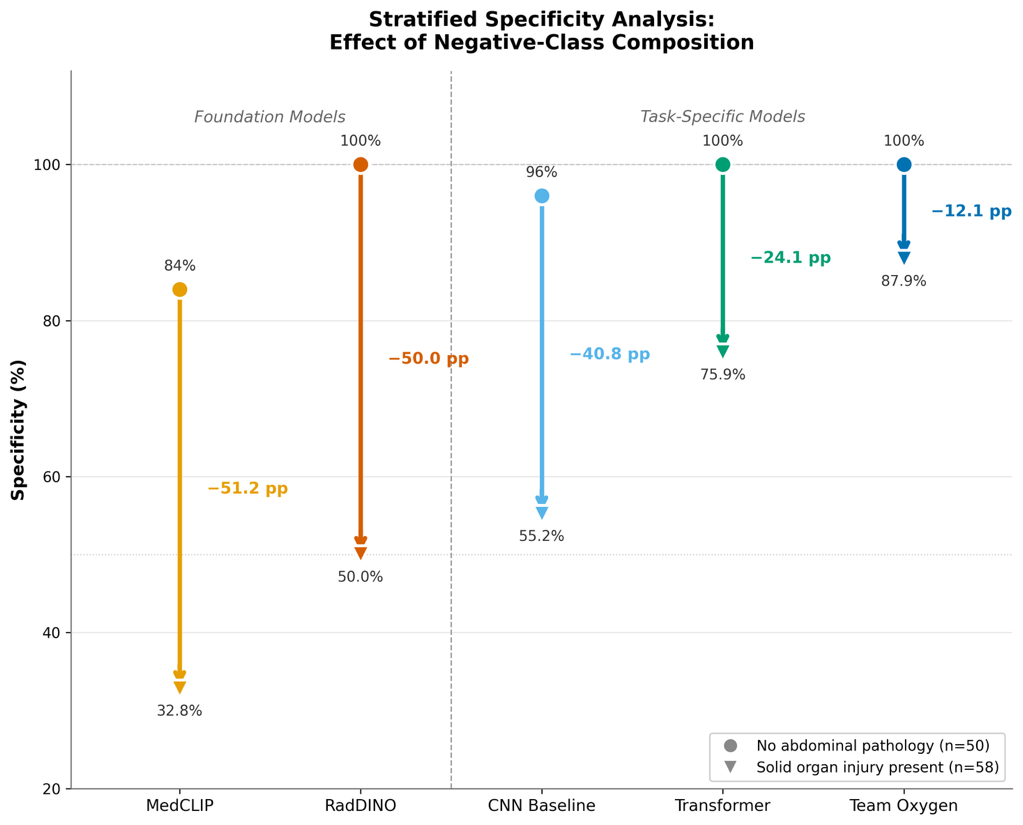}
    \caption{Effect of negative-class composition on model specificity. Each arrow shows the specificity drop from patients without abdominal pathology (circle; n = 50) to patients with solid organ injuries only (triangle; n = 58). Both subgroups have zero bowel injury prevalence. Arrow length indicates magnitude of specificity loss. Models ordered by drop magnitude. Foundation models (warm colours) show 50--51 percentage point drops; task-specific models show 12--41 percentage point drops. LP = linear probe; ZS = zero-shot.}
    \label{fig:stratified}
\end{figure}

To examine whether specificity deficits were associated with negative-class composition, we compared specificity across two negative subgroups with identical bowel injury prevalence (zero): patients without abdominal pathology (n=50) and patients with solid organ injuries only (n=58) (Table \ref{tab:stratified}, Figure \ref{fig:stratified}).

\begin{table}[htbp]
\centering
\caption{Stratified specificity analysis. Specificity was compared between two populations with zero bowel injury prevalence: patients with no abdominal pathology (No pathology) and patients with solid organ injuries but no bowel injury (SOI Present).}
\label{tab:stratified}
\small
\begin{tabular}{llccc}
\toprule
\textbf{Model} & \textbf{Type} & \textbf{No Pathology (n=50)} & \textbf{SOI Present (n=58)} & \textbf{Difference, pp} \\
\midrule
\multicolumn{5}{l}{\textit{Foundation models}} \\
RadDINO & Foundation (LP) & 100.0\% & 50.0\% & $-$50.0 \\
MedCLIP & Foundation (ZS) & 84.0\% & 32.8\% & $-$51.2 \\
\midrule
\multicolumn{5}{l}{\textit{Task-specific models}} \\
CNN Baseline & Task-specific & 96.0\% & 55.2\% & $-$40.8 \\
Team Oxygen & Task-specific & 100.0\% & 87.9\% & $-$12.1 \\
Transformer & Task-specific & 100.0\% & 75.9\% & $-$24.1 \\
\bottomrule
\end{tabular}
\end{table}

Specificity on patients without abdominal pathology was high across all models: RadDINO 100\%, CNN 96.0\%, Transformer 100\%, Team Oxygen 100\%, and MedCLIP 84.0\%. When solid organ injuries were present, foundation model specificity dropped substantially: RadDINO to 50.0\%, MedCLIP to 32.8\%---differences of 50.0 and 51.2 percentage points, respectively. Task-specific models showed smaller differences: CNN 40.8 percentage points (96.0\% to 55.2\%), Transformer 24.1 percentage points (100\% to 75.9\%), and Team Oxygen 12.1 percentage points (100\% to 87.9\%).

This pattern---specificity deficits occurring predominantly when solid organ injuries were present---suggests that organ confusion, whereby models misclassify solid organ injuries as bowel injury, is a major contributor to foundation model specificity deficits. Task-specific training progressively reduced susceptibility to this negative-class heterogeneity.

\subsection{Qualitative Examples}

\begin{figure}[htbp]
    \centering
    \includegraphics[width=0.95\textwidth]{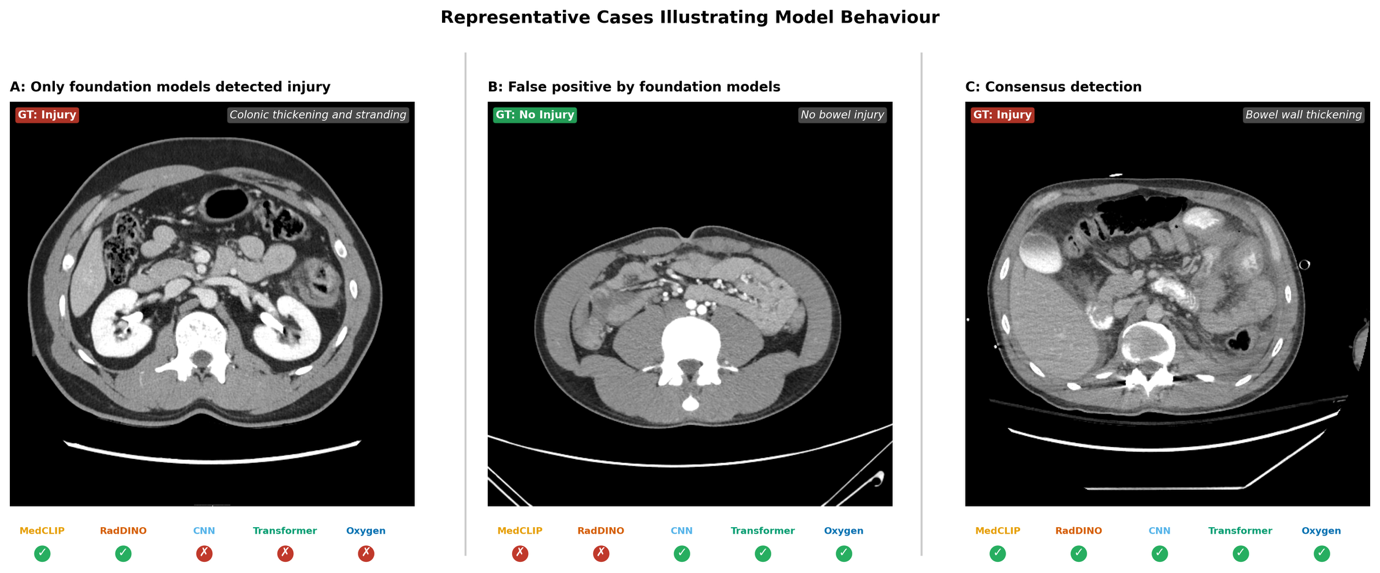}
    \caption{Representative cases illustrating model behavior. (A) Colonic thickening with fat stranding detected only by foundation models (MedCLIP and RadDINO positive; all task-specific models negative), demonstrating sensitivity to heterogeneous presentations. (B) False positive by both foundation models in a patient with liver injury but no bowel injury, illustrating organ confusion. (C) Bowel wall thickening detected by all models (consensus detection). Arrow indicates area of suspected bowel injury.}
    \label{fig:examples}
\end{figure}

Figure \ref{fig:examples} presents representative cases illustrating these patterns. Case A shows colonic thickening with fat stranding detected only by foundation models, demonstrating their sensitivity to heterogeneous presentations. Case B shows a false positive by both foundation models in a patient with liver injury but no bowel injury, illustrating organ confusion. Case C shows bowel wall thickening detected by all models.

\clearpage

\section{Discussion}

This study evaluated foundation models against task-specific approaches for bowel injury detection under compound distribution shift defined as class imbalance co-occurring with appearance heterogeneity. Foundation models achieved equivalent discrimination (AUC 0.64--0.68 versus 0.58--0.64) despite requiring no task-specific training but exhibited lower specificity (33--50\% versus 50--88\%). Stratified analysis revealed that this specificity deficit was strongly associated with negative-class heterogeneity --- specificity dropped 50--51 percentage points when solid organ injuries were present compared to patients without abdominal pathology. These results suggest that confounding pathology, not prevalence miscalibration alone, is the primary driver of foundation model specificity failure in this setting.

The underlying mechanism is that foundation models, through diverse pretraining, develop sensitivity to generic imaging abnormality, for example tissue disruption, free fluid, inflammatory change, but without task-specific supervision, cannot distinguish which organ system is affected \cite{moor2023,denner2024}. We term this organ confusion. Solid organ injuries produce imaging features that overlap substantially with bowel injury: hepatic lacerations generate free fluid, splenic injuries cause tissue heterogeneity, and renal trauma creates retroperitoneal changes \cite{bates2017}. Foundation models flag these findings as suspicious because they are abnormal, but lack the organ-specific discrimination to attribute them correctly. This framing generalises to any scenario where confounding conditions share imaging features with target pathology: chronic microvascular changes mistaken for acute infarcts, post-treatment changes mimicking cancer recurrence, or inflammatory conditions confused with malignancy.

These findings have direct clinical implications. Projecting to clinical prevalence (2.3\%), foundation model performance would yield PPV of approximately 3\% but negative predictive value (NPV) exceeding 99\%, suggesting potential utility as a rule-out tool requiring prospective validation. More broadly, the identification of organ confusion as the primary failure mode provides actionable guidance: effective adaptation must address organ discrimination specifically, not simply recalibrate for prevalence. The stratified analysis methodology introduced here --- comparing performance across negative subgroups with controlled target prevalence --- provides a general framework for diagnosing specificity deficits. It is applicable to any rare disease detection task where hard negatives co-occur with the target pathology.

The specificity problem we identify is not unique to trauma. Across bowel pathology, AI models have achieved strong discrimination for conditions ranging from obstruction and ischaemia to colorectal cancer and inflammatory bowel disease \cite{wang2026,meng2022}. Yet specificity is consistently the weaker metric. Models for bowel obstruction on CT achieved only 76\% specificity on external validation when calibrated for high sensitivity \cite{vanderbecq2024a}. For ischaemia detection --- where findings such as pneumatosis and pneumoperitoneum must be distinguished from mimics --- external specificity fell to 44\% despite 89\% sensitivity \cite{vanderbecq2025}. In each case, false positives were reported but not investigated. No study examined whether the composition of the negative class contributed to these failures, and no study applied a vision-language foundation model to any bowel pathology task. Our work addresses both gaps: it is the first evaluation of a vision-language foundation model for bowel injury detection, and the first to demonstrate that co-occurring pathology among negative cases is a measurable and substantial source of specificity failure.

Foundation model evaluations in medical imaging have addressed prevalence shift and appearance heterogeneity as separate challenges \cite{guo2017,ganin2015,khan2025}, leaving their interaction untested. This matters because foundation models, through diverse pretraining, develop strong sensitivity to generic imaging abnormality, detecting it reliably across modalities and pathologies without task-specific training \cite{moor2023}. However, this generic sensitivity retrieves anatomical features far more accurately than pathological ones \cite{denner2024}, meaning models recognise that something is abnormal more readily than they identify what is abnormal. Our results show that under compound shift, this becomes the organ confusion pattern described above --- overlapping imaging features between bowel and solid organ injuries prevent correct attribution. Both MedCLIP and RadDINO were pretrained predominantly on chest imaging, meaning bowel injury is likely out-of-distribution entirely. Evaluation frameworks that address prevalence shift or appearance heterogeneity in isolation will not detect this failure mode; benchmarks incorporating compound shift with heterogeneous negative classes are needed.

For traumatic bowel injury specifically, the published AI literature is thin. The only prior deep learning study reported specificity of 94.3\% but detected fewer than one in six injuries (sensitivity 14.9\%) \cite{shen2024}. The RSNA RATIC competition achieved mean AUC of 0.85 for bowel and mesenteric injury across eight winning models, but reported aggregate performance only, without stratifying by co-occurring injuries \cite{hermans2025}. Neither study examined how concurrent solid organ injuries --- present in the majority of blunt abdominal trauma patients --- affect bowel injury detection. Our stratified analysis addresses this directly. Because bowel injury prevalence was zero in both negative subgroups, the 50-percentage-point specificity difference between patients with and without solid organ injuries cannot reflect prevalence miscalibration. Foundation models showed similar susceptibility regardless of adaptation level (50--51 percentage point drops), while task-specific models showed progressively smaller reductions with increasing supervision --- from 41 percentage points (CNN) to 24 (Transformer) to 12 (ensemble). This pattern suggests that labelled training data, rather than architectural complexity alone, reduces organ confusion, though the comparison is observational given the architectural differences between models.

The practical appeal of foundation models for rare pathology detection lies in eliminating the need for large, labelled datasets. For traumatic bowel injury, where prevalence is 2--5\% and labelled datasets are scarce, both MedCLIP's zero-shot inference and RadDINO's linear probe achieved discrimination equivalent to fully supervised models --- demonstrating the value of diverse pre-training. However, this comes at a specificity cost when confounding pathology is present. Parameter-efficient approaches such as adapter modules \cite{hu2021} or mixture-of-experts architectures \cite{mu2025} may reduce organ confusion while preserving pre-trained representations, and represent a natural next step toward clinical translation.

This study has several limitations. We evaluated two of the most widely used foundation models; other architectures may behave differently. RadDINO's feature-level max-pooling may dilute injury signal across large volumes, unlike MedCLIP's inherent per-slice classification, a limitation addressable through attention-based aggregation strategies. Neither foundation model processes full 3D volumes natively; both rely on 2D or 2.5D representations that may miss volumetric spatial relationships relevant to bowel injury detection. The enriched test set enabled statistical evaluation but differs from deployment prevalence. The cross-model comparison is observational. Nevertheless, the stratified analysis---enabled by a no-pathology cohort of truly normal patients---controls for prevalence and isolates the effect of negative-class composition, addressing a limitation common to trauma imaging datasets where negative cases typically have concurrent injuries.

\clearpage

\section{Conclusion}

Traumatic bowel injury detection provided a clinical testbed for an unsolved problem in the deployment of AI: how foundation models behave when class imbalance and appearance heterogeneity co-occur. Foundation models matched task-specific discrimination without training, but specificity dropped 50 percentage points when confounding solid organ injuries were present --- revealing that heterogeneity in the negative class, not prevalence alone, drives specificity failure. Task-specific supervision progressively reduced this susceptibility, from zero-shot to fully supervised approaches. Before foundation models can support clinical decision-making, adaptation methods must specifically address the composition of the negative class, a challenge likely to arise wherever confounding pathology co-occurs with the target condition.

\clearpage

\section{Declarations}

\textbf{Competing Interests:} None

\textbf{Funding:} None

\textbf{Ethics Approval:} This study used publicly available, de-identified data from the RSNA RATIC 2024 Challenge. Institutional review board approval was not required.

\textbf{Data Availability:} The RSNA RATIC 2024 dataset is available through the Radiological Society of North America. Code will be made available upon publication.

\end{document}